\documentclass{elsart}

\usepackage{natbib,graphicx,epsfig}

\usepackage{amssymb}

\begin{document}

\begin{frontmatter}



\title{Search for dense molecular gas in QSO hosts}


\author[label1,label4]{J.-U. Pott$^{a,b}$, A. Eckart$^a$, M. Krips$^c$, L. Tacconi-Garman$^b$, and E. Lindt}
\address[label1]{I. Physikalisches Institut, Universit\"at zu K\"oln, 50937 K\"oln, Germany}
\address[label2]{European Southern Observatory, 85748 Garching b. M\"unchen, Germany}
\address[label3]{Smithsonian Astrophysical Observatory, Submillimeter Array, HI 96720, USA}
\address[label4]{Institut de Radio Astronomie Millim\'etrique, 38406 St. Martin d'H\`eres, France}

\begin{abstract}
We present the results of a recently conducted mm-experiment at the IRAM 30m telescope. We searched in two QSO hosts for the HCN(J=1-0) line emission tracing the dense regions of gas. Our goal is to probe the HCN-CO line ratio in QSO hosts for the first time and to compare the results with recent findings of quiescent galaxies. These findings indicate a strong correlation between the star formation efficiency and the HCN-CO line ratio over several orders of magnitudes in infrared luminosities.
\end{abstract}


\end{frontmatter}

\section{Introduction}
\label{sec:intro}
The evolutionary sequence from ULIRGs to quasars (Sanders 1999, \& Sanders 2003 for reviews) has become a popular, though controversial, theory. Numerous publications deal with the AGN components and the merger morphologies at the ULIRG stage. But the starburst properties and merger signs of post-ULIRG quasars have hardly been addressed. 
\newline
Evans et al. (2001) investigated the total molecular gas content of infrared-bright, optically selected QSOs. These galaxies may evolve out of ULIRGs and show systematically higher infrared-to-CO luminosity ratios compared to the majority of apparently quiescent galaxies (i.e. with no obvious AGN) with similarly high infrared luminosity. This lead Evans et al. (2001) to the interpretation that their selected sources may show a strong AGN heated dust contribution to the infrared luminosity. An alternative explanation would be an increase in star formation efficiency, i.e. star formation rate with respect to the total amount of molecular gas available. The latter is traced by the CO rotational line transitions.\newline
To test which scenario is more important an independent check of the star formation efficiency has to be performed. Gao \& Solomon (2004; GS04) investigated a large sample of 65 nearby galaxies, covering three orders of magnitude of infrared luminosity L$_{IR}$ up to the ULIRG regime (L$_{IR}$ $> 10^{12}$ L$_\odot$). They studied the HCN content and established that the {\em dense} gas tracer HCN is tightly correlated to the SFR by L$_{IR}$/L$_{HCN}$=900~L$_\odot$(K km s$^{-1}$pc$^2$)$^{-1}$. A similar correlation between SFR as traced by L$_{IR}$ and the {\em total} molecular gas content traced by L$_{CO}$ breaks down towards the high infrared luminosities of ULIRGs (L$_{IR}$/L$_{CO}$=33~L$_\odot$(K km s$^{-1}$pc$^2$)$^{-1}$). GS04 concluded in their study that the global HCN/CO line intensity ratio is a strong starburst indicator.
\newline
By applying this indicator to their sample the total infrared luminosity could be assigned to starburst activity, excluding AGN heated dust. This holds for the entire sample, including the most infrared luminous ULIRGs. \newline
Thus the ULIRG phenomenon in nearby galaxies without indications for nuclear activity from other wavelength regimes appears to result exclusively from enhanced star formation efficiency (SFE). The question of AGN contribution to  L$_{IR}$ remains open for infrared bright QSO+hostgalaxy systems, which are shown to harbour an AGN in the optical wavelength regime.
\newline To combine the findings of Evans et al. (2001) and GS04, we have observed the two CO-brightest infrared-excess QSOs (PG0838+770 and PG1415+451) of the sample of Evans et al. (2001) in the rotational line transition of HCN(J=1-0) to trace the star formation efficiency and star burst contribution to the infrared luminosity in these galaxies. Both sources have IR luminosities in the LIRG range (10$^{11} \le $ L$_\odot$ $\le$ 10$^{12}$). Haas et al. 2003 have classified them within a wide-range spectral energy distribution survey to be evolved quasars between the ULIRG and old quasars stage. 
Thus the selected targets are ideal to complement the normal-to-(U)LIRG findings of GS04 with similar observations of quasar hosts.

\section{Observations and Results}
\label{sec:obs}

   \begin{figure*}
   \centering
   \includegraphics[width=12cm]{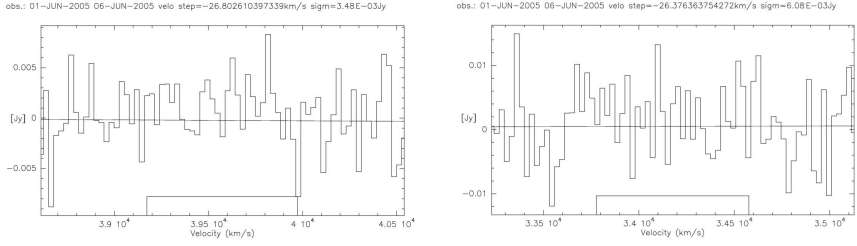}
   \caption{HCN in PG0838+770 (left):1$\sigma$=3.5~mJy and PG1415+451 (right): 1$\sigma$=6~mJy}
              \label{fig:1}%
    \end{figure*}

The observations were carried out in June 2005 using the IRAM 30 meter
radiotelescope on Pico Veleta, near Granada, Spain.
Weather conditions were good during the run with typical
system temperatures of 150-250 K and the 3mm opacity at zenith was
$0.05 \le \tau_{zen} \le 0.1$.  Because the (redshifted) observing frequencies fall outside the standard tuning range of the
receivers, we made use of the recently offered workaround requiring a
non-standard tuning of the 3mm receivers with the 1mm local
oscillators. A freqency tuning test on the giant molecular cloud Sgr B2 was successfully performed. More details about the observational setup used will be published soon (Pott et al., in prep.)\newline
Our observed spectra and the signal-to-noise ratio achieved are shown in Fig.\ref{fig:1}.
We did not detect any HCN(J=1-0) line emission in both galaxies at the given noise level. Because the redshifts have been confirmed spectroscopically by the preceding CO observations, we can derive upper limits on basis of the CO findings. Assuming a linewidth similar to that of CO we derive the following 3$\sigma$ upper limits to the luminosity of the HCN(J=1-0) rotational line transition:
$$ L'_{{\rm HCN}}({\rm 0838})\,\le \,10\cdot 10^8\,{\rm K\,km\,s^{-1}\,pc^2}; L'_{{\rm HCN}}({\rm 1415})\,\le \,\,7 \cdot 10^8\,{\rm K\,km\,s^{-1}\,pc^2}$$
   \begin{figure*}
   \centering
   \includegraphics[height=5cm]{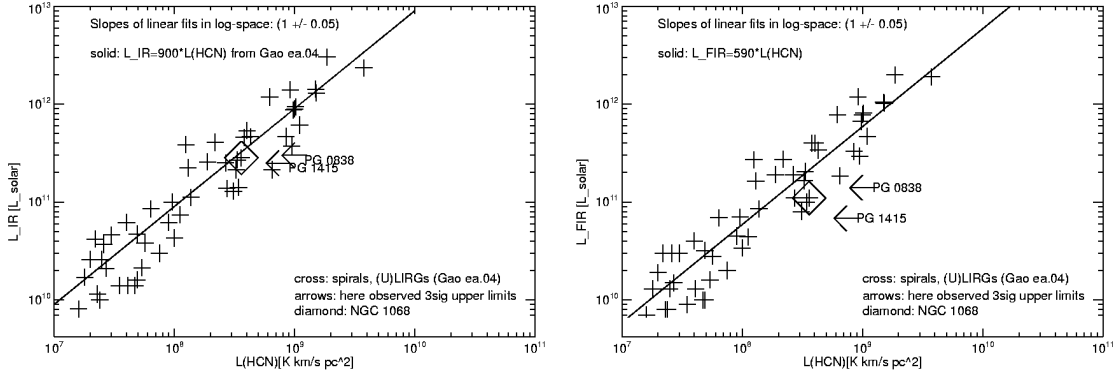}
   \caption{Here our upper limits (arrows) are confronted with the data published in GS04. The definition of L$_{\rm IR}$ and L$_{\rm FIR}$ can be found e.g. in Sanders et al. 2003 and references therein. The size of the crosses is of the order of the uncertainties. }
              \label{fig:2}%
    \end{figure*}

   \begin{figure*}
   \centering
   \includegraphics[height=4cm]{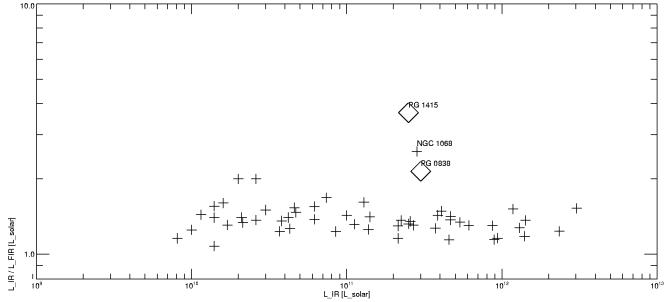}
   \caption{This luminosity ratio indicates an additional AGN heated near-infrared component in the observed quasar hosts, in contrast to the majority of the nearby quiescent quasars, including ULIRGs. NGC~1068, known to bear an AGN, lies between the observed quasars.}
              \label{fig:3}%
    \end{figure*}

\section{Summary}
\label{sec:sum}
In Fig.\ref{fig:2} these luminosity limits are put in the context of the galaxy sample, observed by GS04. Our upper limits (as illustrated by arrows) indicate general agreement with the results of the reference sample. With the published CO luminosities we have calculated an upper limit of the HCN(1-0)-to-CO(1-0) luminosity ratio of about 0.5 for the observed quasar hosts. Similar investigations in nearby galaxies found ratios around 0.05 for normal galaxies and values increased up to 0.3 in ULIRGs and in the centers of Seyferts (GS04, Kohno et al. 2003). Recent investigations revealed abnormally high {\em local} ratios larger than 0.5 very close to AGNs (e.g. NGC~1068; Usero et al. 2004), suggesting that here the HCN is excited by the AGN emission rather than by starbursts. Assuming similar excitation mechanisms close to a QSO, we can exclude, that the AGN in our observed QSOs is dominating the whole galaxy in a way, that {\em global} HCN/CO ratios above 0.5 appear. 
\newline
In Fig.\ref{fig:3} we plotted the IR-to-FIR luminosity ratio of the investigated galaxies. Here the distinct ratio seen for the quasar hosts is obvious. Only the known active galaxy of Seyfert II type, NGC~1068, shows similar flux ratios. This suggests, that the complete IR luminosity (8-1000~$\mu$m) of quasar hosts comprises a significant contribution of hot dust which is not apparent in non-active galaxies, even in the ULIRG stage. The central AGN is most probably responsible for that contribution. Similarly the {\em global}  L$_{\rm FIR}$-HCN ratio in NGC~1068 seems to be smaller than usual (Fig.\ref{fig:2}), indicating an additional HCN excitation by the AGN. This has an impact on the design of future studies of quasar hosts. If the HCN/CO ratio is indeed a good tracer for the star formation efficiency, the HCN luminosity should be tightly correlated only to the L$_{\rm FIR}$ in quasar host, whereas the AGN contribution to the L$_{\rm IR}$ breaks up a similar correlation. On the other hand an additional excitation by the AGN would favour a break up of the L$_{\rm FIR}$-HCN correlation.
In the near future we will broaden our sample to increase the statistical significance of our findings.




\end{document}